\documentclass[journal]{IEEEtran}
\usepackage[OT1]{fontenc}
\usepackage[utf8]{inputenc}
\usepackage{xcolor}
\usepackage{bm}
\usepackage{amsmath}
\usepackage{amssymb}
\usepackage{graphicx}
\PassOptionsToPackage{normalem}{ulem}
\usepackage{ulem}

\makeatletter
\allowdisplaybreaks

\IEEEoverridecommandlockouts

\usepackage{color}
\usepackage{bm}
\usepackage{algorithm}
\usepackage{algorithmic}
\usepackage{multirow}
\usepackage{booktabs}
\usepackage{array}
\usepackage{amsthm}
\usepackage{lipsum}
\usepackage{enumerate}
\usepackage{stfloats}
\usepackage{subfigure}
\usepackage{cases}
\usepackage{diagbox}
\usepackage{threeparttable}
\usepackage{cite}
\usepackage{url}

\title{Antenna Coding Design for Multi-User Transmissions Using Pixel Antennas}

\author{Hongyu Li, \textit{Member,~IEEE} and Shanpu Shen, \textit{Senior Member,~IEEE}
\thanks{Manuscript received; This work was funded by the Science and Technology Development Fund, Macau SAR (File/Project no. 001/2024/SKL). Part of this paper has been published and presented in IEEE Workshop on Signal Processing and Artificial Intelligence for Wireless Communications (SPAWC), 2025 \cite{spawc2025}. \textit{(Corresponding author: Shanpu Shen.)}}

\thanks{H. Li is with the Internet of Things Thrust, The Hong Kong University of Science and Technology (Guangzhou), Guangzhou 511400, China (e-mail: hongyuli@hkust-gz.edu.cn).}
\thanks{S. Shen is with the State Key Laboratory of Internet of Things for Smart City and Department of Electrical and Computer Engineering, University of Macau, Macau, China (email: shanpushen@um.edu.mo).}}

\makeatother

\begin{document}
\maketitle \thispagestyle{empty}
\begin{abstract}
This work investigates exploiting the potential of pixel antennas,
which are a reconfigurable antenna technology that can flexibly adjust
the antenna characteristics through antenna coding, in multi-user
transmissions. To that end, we propose a multi-user multi-input single-output
(MISO) pixel antenna system, which deploys the pixel antenna at users,
and develop the system model including pixel antenna with antenna
coding and multi-user beamspace channels. Aiming at maximizing the
sum rate performance, we first propose an algorithm to alternatively
design the precoding at the transmitter and the antenna coding at
users, which explores the performance boundary for the proposed multi-user
MISO pixel antenna system. To reduce the computational complexity,
we propose a codebook-based antenna coding design algorithm, where
the antenna coder is online optimized from an offline codebook. To
further enhance the computation efficiency, we propose a hierarchical
codebook-based antenna coding design that uses a multi-layer hierarchical
search to achieve a better performance-complexity trade-off. Simulation
results show that, adopting the proposed algorithms, the multi-user
MISO pixel antenna system can always outperform conventional multi-user
MISO systems with fixed antennas. More importantly, results validate
that the proposed (hierarchical) codebook-based algorithms can significantly
reduce the computational complexity while maintaining a satisfactory
sum rate performance.
\end{abstract}


\section{Introduction}

Multiple-input multiple-output (MIMO), supported by multiple antennas
at transceivers, has been one of the most dominant and successful
technologies in wireless communications to enable multiple access,
higher throughput, and enhanced reliability \cite{lu2014overview}.
In conventional MIMO systems, antennas usually have fixed configurations
and characteristics, such as radiation pattern and polarization. Therefore,
while sophisticated signal processing techniques have been developed
to adapt to the wireless propagation environment, antennas are excluded
from the system optimization in conventional MIMO communications,
which limits the performance of wireless communication systems. In
this sense, it is necessary to explore the degrees of freedom that
antennas can provide through making the antenna configurations and
characteristics reconfigurable.

Pixel antennas, which is a highly reconfigurable antenna technology
to dynamically reconfigure antenna characteristics, give a promising
solution to break through the performance limitation of conventional
MIMO communication systems \cite{cetiner2004multifunctional,pringle2004reconfigurable}.
The concept of the pixel antenna is based on discretizing a continuous
radiation surface into a grid of small elements called pixels and
interconnecting them with RF switches \cite{chiu2012frequency,song2013efficient}.
Dynamically turning on or off RF switches can control connecting or
disconnecting adjacent pixels to electrically adjust the antenna topology
in real-time, which subsequently allows to reconfigure the antenna
characteristics such as operating frequency \cite{song2013efficient}
and radiation patterns \cite{zhang2022highly,zhang2022new}
to achieve higher flexibility in frequency and beam control. These
appealing advantages of pixel antennas have also motivated different
applications, such as implementing reconfigurable intelligent surfaces
with pixelated elements \cite{rao2022novel}, pixel rectennas to improve
RF energy harvesting \cite{shen2017multiport}, MIMO antennas with
pixelated surface to improve the ergodic capacity \cite{zhang2021compact},
and circular polarization to avoid polarization mismatch \cite{zhang2020polarization}.
Nevertheless, these works \cite{cetiner2004multifunctional,pringle2004reconfigurable,chiu2012frequency,song2013efficient,zhang2022highly,zhang2022new,rao2022novel,shen2017multiport,zhang2021compact,zhang2020polarization} only focus on the antenna
hardware level, overlooking the impact and potential of pixel antennas
on the system level.

In the system level, pixel antennas are related to an emerging technology,
namely fluid antenna systems \cite{wong2020fluid,new2024tutorial,zhu2023movable,zhu2025tutorial}.
The main idea of fluid antenna systems is that the fluid antenna can
freely flow to any desired position within a designated space or aperture,
which can be implemented by mechanically flexible antennas like liquid
antennas using liquid metals \cite{bo2018recent} or alternatively by pixel antennas
that adjust RF switches to mimic switching positions \cite{zhang2024novel}.
Based on the position-switchable property, the fluid antenna can exploit
the spatial variations in a wireless channel to find the position
with the strongest signal, which therefore provides diversity gain
\cite{psomas2023diversity} and enhanced channel gain \cite{zhu2023modeling}.
In addition, fluid antennas have been adopted to construct MIMO systems
with improved capacity, spectral efficiency, and energy efficiency
\cite{new2024tutorial,zhu2025tutorial,new2023information,ma2023mimo}
and also enable a novel multiple access scheme, namely fluid antenna
multiple access \cite{wong2021fluid,wong2022fast,wong2023slow}, which
allow multiple users equipped with fluid antennas to independently
and dynamically reposition the fluid antenna to a location with a
strong signal from their intended transmitter while simultaneously
having weak interference from other users. There have been other applications
of fluid antenna systems, such as secret communications \cite{tang2023fluid},
integrated sensing and communication \cite{zhou2024fluid}, and mobile
edge computing \cite{zuo2024fluid}.

While these works of fluid antenna system \cite{psomas2023diversity,zhu2023modeling,new2023information,ma2023mimo,wong2021fluid,wong2022fast,wong2023slow,tang2023fluid,zhou2024fluid,zuo2024fluid}
have preliminarily disclosed the usefulness of pixel antennas in future
wireless communication systems, to further exploit the significant
potential of pixel antenna in the system level, a recent work \cite{shen2024antenna}
has, for the first time, generalized the radiation pattern reconfigurability
of pixel antenna by deriving a physical and electromagnetic compliant
communication model and proposed a novel antenna coding technique
empowered by pixel antennas to design and enhance wireless communications.
For antenna coding, the antenna coders, which are binary variables
representing the RF switch states, are optimized to reconfigure the
pixel antenna, so as to adjust the radiation pattern and subsequently
enhance the wireless system performance. It has shown that using pixel
antennas to replace conventional antennas with fixed configurations
in point-to-point single-input single-output (SISO) and MIMO systems
can improve the average channel gain by more than 5 times and the
channel capacity by up to 3.1 times, respectively \cite{shen2024antenna}.
In spite of these performance enhancements in the point-to-point wireless
transmission, it remains unexplored whether or not this antenna coding
technique based on pixel antennas can be beneficial in achieving better
interference managements and higher-quality multiple access for multi-user
wireless communication systems.

To address the above concern, in this paper, we consider a multi-user
multi-input single-output (MU-MISO) pixel antenna system where the
transmitter is equipped with conventional antennas with fixed configurations,
while each user is equipped with a pixel antenna. Based on this MU-MISO
pixel antenna system, we optimize the antenna coding to demonstrate
the advantages of pixel antennas in improving sum rate performance
for the multi-user transmission. This leads to the following contributions.

\textit{First}, we establish a tractable yet physical and electromagnetic
compliant communication model for MU-MISO pixel antenna systems, from
which we prove that a MISO channel based on multiple conventional
antennas at the transmitter and single pixel antenna at the user
can be equivalently formulated as a MIMO channel with a pattern coder
controlled by the pixel antenna. This allows us to understand the
benefit of the pixel antenna from the modeling perspective.

\textit{Second}, aiming at maximizing the sum rate performance of
the MU-MISO system using pixel antennas at the user side, we formulate
the joint optimization of the antenna coding at users and the precoding
at transmitter. To solve this joint optimization, we develop an algorithm
to alternatively optimize the antenna coding through successive exhaustive
boolean optimization (SEBO) \cite{shen2016successive} and the precoding through fractional programming \cite{shen2018fractional}.
We also provide the computational complexity analysis for the proposed
algorithm.

\textit{Third}, we propose a joint precoding and codebook-based antenna
coding optimization algorithm to reduce the computational complexity,
where an offline codebook is pre-designed to enable the online optimization
for antenna coding. In this algorithm, the antenna coder for each
user is directly selected from the pre-designed codebook using a one-dimensional
exhaustive search and the corresponding computational complexity analysis
is provided.

\textit{Fourth}, to further reduce the computational complexity, we
design a hierarchical codebook with multiple layers and propose a
joint precoding and hierarchical codebook-based antenna coding optimization
algorithm. In this case, the antenna coder for each user is optimized
by a hierarchical search over layers of the codebook, instead of performing
the one-dimensional exhaustive search over all antenna coders, which
avoids the exhaustive search and further reduces the computational
complexity.

\textit{Fifth}, we provide simulation results to evaluate the performance
of the MU-MISO pixel antenna system. It is demonstrated that replacing
the conventional antenna with pixel antenna at the user side can effectively
improve system performance. For example, the sum rate of the MU-MISO system including two fixed conventional transmit antennas and two users
can be doubled for signal-to-noise-ratio (SNR) equal to 10 dB, by using pixel antennas at each user with optimized antenna coding. 
More importantly, the proposed (hierarchical) codebook-based
algorithms can significantly reduce the computational complexity while
maintaining a good sum rate performance, providing a practical and
efficient solution to apply pixel antennas in wireless systems.

\textit{Organization:} Section \ref{sec:antenna_model} introduces
the antenna coding technique using pixel antennas. Section \ref{sec:system_model}
introduces the channel model of the MU-MISO system using pixel antennas
in beamspace. Section \ref{sec:joint_opt} proposes a joint antenna
coding and precoding design to maximize the sum rate. Section \ref{sec:codebook}
proposes a codebook-based antenna coding design. Section \ref{sec:hierarchical_codebook}
proposes a hierarchical codebook for antenna coding design. Section
\ref{sec:performance} evaluates the performance of the MU-MISO system
using pixel antennas. Section \ref{sec:conclusion} concludes this
work.

\textit{Notations:} $\mathbb{C}$ and $\mathbb{R}$ denote the set
of complex and real numbers, respectively. $(\cdot)^{\mathsf{T}}$,
$(\cdot)^{*}$, $(\cdot)^{\mathsf{H}}$, and $(\cdot)^{-1}$ denote
the transpose, conjugate, Hermitian, and inverse, respectively. $\mathsf{diag}(a_{1},\ldots,a_{M})$
denotes a diagonal matrix with diagonal entries $a_{1},\ldots,a_{M}$.
$\|\cdot\|_{2}$ and $\|\cdot\|_{\mathsf{F}}$ denote the $\ell$-2
norm of a vector and the Frobenius norm of a matrix, respectively.
$|\mathcal{A}|$ denotes the size of a set $\mathcal{A}$. 
$\bigcup_{n=1}^N \mathcal{A}_i$ denotes the union of sets $\mathcal{A}_1,\ldots,\mathcal{A}_N$. 
$\mathbb{E}\{\cdot\}$ denotes the expectation. $\mathbf{I}_{N}$
denotes an $N\times N$ identical matrix. $\mathcal{CN}(0,\sigma^{2})$
denotes the circularly symmetric complex Gaussian distribution with
zero mean and covariance $\sigma^{2}$. $\lceil\cdot\rceil$ denotes
the ceiling function. $\mathsf{mod}(M,N)$ returns the remainder after
$M$ is divided by $N$. $[\mathbf{A}]_{:,i}$ and $[\mathbf{A}]_{i,j}$ denote the $i$th column and the $(i,j)$th entry of a matrix $\mathbf{A}$, respectively. 

\section{Antenna Coding Based on Pixel Antennas}

\label{sec:antenna_model}

As the starting point, in this section we briefly introduce the model
of pixel antenna using multiport network theory and the antenna coding
technique.

\begin{figure}
\centering \includegraphics[width=0.48\textwidth]{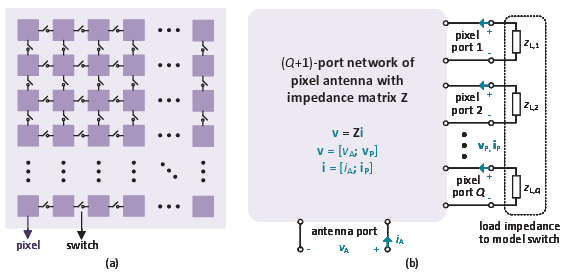}
\caption{(a) Illustration of a pixel antenna. (b) Multiport network model for
pixel antenna (achieved by replacing switches with ports).}
\label{fig:pixel_antenna_model} 
\end{figure}

Pixel antennas are based on discretizing a continuous radiation surface
into small elements called pixels, where adjacent pixels are connected
by RF switches such that the antenna configuration can be flexibly
adjusted to implement a highly reconfigurable antenna \cite{song2013efficient,shen2024antenna},
as illustrated in Fig. \ref{fig:pixel_antenna_model}(a).

Based on multiport network theory, a pixel antenna embedded with $Q$
RF switches can be modeled as a $(Q+1)$-port network consisting of
one antenna port and $Q$ pixel ports, as shown in Fig. \ref{fig:pixel_antenna_model}(b).
The $(Q+1)$-port network is characterized by its impedance matrix
$\mathbf{Z}\in\mathbb{C}^{(Q+1)\times(Q+1)}$. Therefore, we have
$\mathbf{v}=\mathbf{Z}\mathbf{i}$, where the voltage vector $\mathbf{v}=[v_{\mathrm{A}};\mathbf{v}_{\mathrm{P}}]\in\mathbb{C}^{(Q+1)\times1}$
and current vector $\mathbf{i}=[i_{\mathrm{A}};\mathbf{i}_{\mathrm{P}}]\in\mathbb{C}^{(Q+1)\times1}$
collect the voltage $v_{\mathrm{A}}\in\mathbb{C}$ and current $i_{\mathrm{A}}\in\mathbb{C}$
at the antenna port, and voltages $\mathbf{v}_{\mathrm{P}}\in\mathbb{C}^{Q\times1}$
and currents $\mathbf{i}\in\mathbb{C}^{Q\times1}$ at the pixel ports,
respectively. Accordingly, $\mathbf{Z}$ can be partitioned as 
\begin{equation}
\mathbf{Z}=\left[\begin{matrix}z_{\mathrm{AA}} & \mathbf{z}_{\mathrm{AP}}\\
\mathbf{z}_{\mathrm{PA}} & \mathbf{Z}_{\mathrm{PP}}
\end{matrix}\right],
\end{equation}
where $z_{\mathrm{AA}}\in\mathbb{C}$ and $\mathbf{Z}_{\mathrm{PP}}\in\mathbb{C}^{Q\times Q}$
denote the self-impedance (matrix) for the antenna port and pixel
ports, respectively, $\mathbf{z}_{\mathrm{PA}}\in\mathbb{C}^{Q\times1}$
denotes the trans-impedance relating the current of antenna port and
voltages of pixel ports, and $\mathbf{z}_{\mathrm{AP}}=\mathbf{z}_{\mathrm{PA}}^{\mathsf{T}}$.

Meanwhile, the pixel port $q$, $\forall q\in\mathcal{Q}=\{1,\ldots,Q\}$
is connected to load impedance $z_{\mathrm{L},q}$ characterizing
the on and off states of the $q$th switch. That is, when the switch
is turned on, we have $z_{\mathrm{L},q}$ is short-circuit, i.e. $z_{\mathrm{L},q}=0$;
when the switch is turned off, we have $z_{\mathrm{L},q}$ is open-circuit\footnote{The infinite for the open-circuit state of the switch is numerically
approximated by setting $z_{\mathrm{L},q}=\jmath\beta$, where $\jmath$
is the imaginary unit and $\beta\in\mathbb{R}$ has a sufficiently
large value, such as $\beta=10^{10}$.}, i.e. $z_{\mathrm{L},q}=\infty$. This allows us to introduce a binary
vector $\mathbf{b}=[b_{1},\ldots,b_{Q}]^{\mathsf{T}}\in\{0,1\}^{Q\times1}$,
which is called the antenna coder, to represent the states of $Q$
switches and yields the diagonal load impedance matrix $\mathbf{Z}_{\mathrm{L}}(\mathbf{b})=\mathsf{diag}(z_{\mathrm{L},1},\ldots,z_{\mathrm{L},Q})\in\mathbb{C}^{Q\times Q}$
with 
\begin{equation}
z_{\mathrm{L},q}=\begin{cases}
\infty, & \text{if }b_{q}=1,\\
0, & \text{if }b_{q}=0.
\end{cases}
\end{equation}
The load impedance matrix relates the voltage $\mathbf{v}_{\mathrm{P}}$
and current $\mathbf{i}_{\mathrm{P}}$ by $\mathbf{v}_{\mathrm{P}}=-\mathbf{Z}_{\mathrm{L}}(\mathbf{b})\mathbf{i}_{\mathrm{P}}$.
As such, the currents at the pixel ports can be related to the current
at the antenna port through 
\begin{equation}
\mathbf{i}_{\mathrm{P}}(\mathbf{b})=-(\mathbf{Z}_{\mathrm{PP}}+\mathbf{Z}_{\mathrm{L}}(\mathbf{b}))^{-1}\mathbf{z}_{\mathrm{PA}}i_{\mathrm{A}},
\end{equation}
which is coded by the antenna coder $\mathbf{b}$.

With the coded currents $\mathbf{i}_{\mathrm{P}}(\mathbf{b})$ at
the pixel ports, we have the corresponding coded radiation pattern
of the pixel antenna, which is the superposition of the radiation
patterns at all $Q+1$ ports weighted by the currents $\mathbf{i}(\mathbf{b})=[i_{\mathrm{A}};\mathbf{i}_{\mathrm{P}}(\mathbf{b})]$,
that is 
\begin{equation}
\mathbf{e}(\mathbf{b})=\mathbf{E}_{\mathrm{oc}}\mathbf{i}(\mathbf{b}),\label{eq: e=00003DEoci}
\end{equation}
where $\mathbf{E}_{\mathrm{oc}}=[\mathbf{e}_{\mathrm{A}}^{\mathrm{oc}},\mathbf{e}_{\mathrm{P},1}^{\mathrm{oc}},\ldots,\mathbf{e}_{\mathrm{P},Q}^{\mathrm{oc}}]\in\mathbb{C}^{2K\times(Q+1)}$
collects the open-circuit radiation pattern\footnote{Here, the open-circuit radiation pattern refers to the radiation pattern
excited by unit current at one port with all the other ports open-circuit.} $\mathbf{e}_{\mathrm{A}}^{\mathrm{oc}}\in\mathbb{C}^{2K\times1}$
at the antenna port (including azimuth and elevation polarization
components over $K$ spatial angle samples) and the open-circuit radiation
patterns $\mathbf{e}_{\mathrm{P},q}^{\mathrm{oc}}\in\mathbb{C}^{2K\times1}$,
$\forall q\in\mathcal{Q}$, at $Q$ pixel ports. Thus, by flexibly
selecting the antenna coder $\mathbf{b}$ among the $2^{Q}$ different
combinations, the radiation pattern of the pixel antenna $\mathbf{e}(\mathbf{b})$
can be effectively reconfigured, which provides additional degrees
of freedom to design and enhance wireless systems.

\section{MU-MISO System Using Pixel Antennas}

\label{sec:system_model}

\begin{figure}
\centering \includegraphics[width=0.48\textwidth]{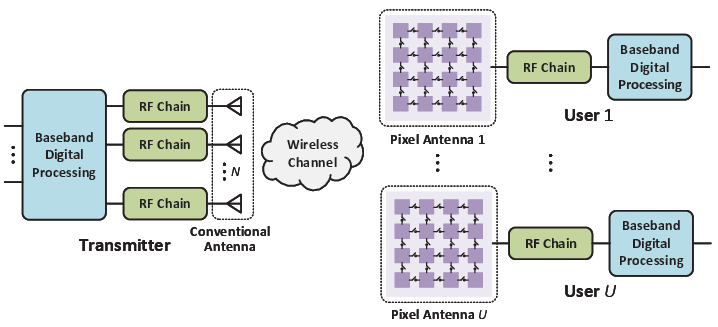}
\caption{Diagram of a multi-user MISO system consisting of an $N$-antenna
transmitter with conventional antennas and $U$ users, each of which
is equipped with a pixel antenna.}
\label{fig:system_model} 
\end{figure}

To show the benefits of pixel antennas to enhance wireless communication
systems, we consider a typical MU-MISO system consisting of a transmitter
with $N$ conventional antennas\footnote{Here, conventional antennas refer to antennas which have a fixed configuration
and radiation pattern.} and $U$ users. Each user is equipped with a single pixel antenna,
as illustrated in Fig. \ref{fig:system_model}. Denote the symbol
vector transmitted to $U$ users as $\mathbf{s}=[s_{1},\ldots,s_{U}]^{\mathsf{T}}$
with $\mathbb{E}\{\mathbf{s}\mathbf{s}^{\mathsf{H}}\}=\mathbf{I}_{U}$.
The transmit symbol $\mathbf{s}$ is precoded by a precoder matrix
$\mathbf{P}=[\mathbf{p}_{1},\ldots,\mathbf{p}_{U}]\in\mathbb{C}^{N\times U}$,
yielding the transmit signal $\mathbf{x}=\mathbf{P}\mathbf{s}$. The
precoder matrix is constrained by a power constraint, i.e. $\|\mathbf{P}\|_{\mathsf{F}}^{2}\le P$,
where $P$ denotes the total transmit power. In the sequel, we will
introduce the beamspace channel representation for pixel antennas,
followed by the beamspace channel reformulation to make it more tractable.

\subsection{Beamspace Channel for Pixel Antennas}

Representing a wireless channel in beamspace \cite{pillai2012array,kalis2014parasitic}
has its unique advantage in analyzing the impact of various antenna
characteristics, such as the radiation pattern, on the wireless channel.

\begin{figure}
\centering \includegraphics[width=0.48\textwidth]{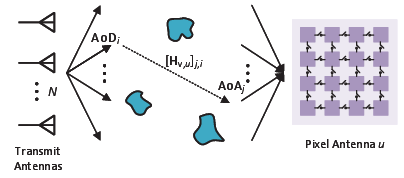}
\caption{Illustration of the beamspace channel representation for pixel antennas. }
\label{fig:channel_model}
\end{figure}

As illustrated in Fig. \ref{fig:channel_model}, for the considered
MU-MISO system, denoting the radiation pattern of the $n$th transmit
antenna as $\mathbf{e}_{\mathrm{T},n}$ $\forall n=1,\ldots,N$ and
the radiation pattern of the pixel antenna at user $u$ as $\mathbf{e}_{u}(\mathbf{b}_{u})$
where $\mathbf{b}_{u}$ is the antenna coder for user $u$, we can
formulate the beamspace MISO channel as 
\begin{equation}
\mathbf{h}_{u}(\mathbf{b}_{u})=\mathbf{e}_{u}^{\mathsf{T}}(\mathbf{b}_{u})\mathbf{H}_{\mathrm{v},u}\mathbf{E}_{\mathrm{T}},\forall u,\label{eq: beamspace channel}
\end{equation}
where $\mathbf{E}_{\mathrm{T}}=[\mathbf{e}_{\mathrm{T},1},\ldots,\mathbf{e}_{\mathrm{T},N}]\in\mathbb{C}^{2K\times N}$
collects the radiation patterns of all transmit antennas and satisfies
that $\mathbf{E}_{\mathrm{T}}^{\mathsf{H}}\mathbf{E}_{\mathrm{T}}=\mathbf{I}_{N}$
since we assume the $N$ transmit antennas are spatially separated
with orthogonal radiation patterns, and $\mathbf{H}_{\mathrm{v},u}\in\mathbb{C}^{2K\times2K}$
denotes the virtual channel matrix for user $u$ given by
\begin{equation}
\mathbf{H}_{\mathrm{v},u}=\left[\begin{array}{cc}
\mathbf{H}_{\mathrm{v},\theta\theta,u} & \mathbf{H}_{\mathrm{v},\theta\phi,u}\\
\mathbf{H}_{\mathrm{v},\phi\theta,u} & \mathbf{H}_{\mathrm{v},\phi\phi,u}
\end{array}\right],\label{eq:virtual channel matrix}
\end{equation}
where $\mathbf{H}_{\mathrm{v},\theta\theta,u}$, $\mathbf{H}_{\mathrm{v},\theta\phi,u}$,
$\mathbf{H}_{\mathrm{v},\phi\theta,u}$, $\mathbf{H}_{\mathrm{v},\phi\phi,u}\in\mathbb{C}^{K\times K}$
are the virtual channel matrices for elevation $\theta$ and azimuth
$\phi$ polarizations for user $u$, respectively, with each entry
representing the channel gain from an angle of departure (AoD) to
an angle of arrival (AoA) among the $K$ spatial angle samples. Note
that the radiation patterns for transmit antennas and pixel antennas
at all users are normalized, i.e. $\|\mathbf{e}_{\mathrm{T},n}\|_{2}=1$
$\forall n=1,\ldots,N$ and $\|\mathbf{e}_{u}(\mathbf{b}_{u})\|_{2}=1$
$\forall u\in\mathcal{U}$, to ensure the normalized channel gain.
From the beamspace channel representation \eqref{eq: beamspace channel},
we can find that the radiation pattern $\mathbf{e}_{u}(\mathbf{b}_{u})$ for user $u$ is reconfigurable
and explicitly controlled by the antenna coder $\mathbf{b}_{u}$,
which allows us to optimize the antenna coder to adapt to the channel
to enhance the wireless system.

\subsection{Beamspace Channel Reformulation}

To simplify the beamspace channel representation, making use of \eqref{eq: e=00003DEoci},
we can rewrite \eqref{eq: beamspace channel} as
\begin{equation}
\mathbf{h}_{u}(\mathbf{b}_{u})=\mathbf{i}_{u}^{\mathsf{T}}(\mathbf{b}_{u})\mathbf{E}_{\mathrm{oc}}^{\mathsf{T}}\mathbf{H}_{\mathrm{v},u}\mathbf{E}_{\mathrm{T}},\forall u,\label{eq: beamspace channel i}
\end{equation}
where we assume that the pixel antennas at $U$ users are identical
so that they have the same open-circuit radiation pattern matrix $\mathbf{E}_{\mathrm{oc}}$.
Accordingly, we perform singular value decomposition for the open-circuit
radiation pattern matrix as
\begin{equation}
\mathbf{E}_{\mathrm{oc}}=\mathbf{U}\mathbf{S}\mathbf{V}^{\mathsf{H}},\label{eq:SVD}
\end{equation}
where $\mathbf{U}\in\mathbb{C}^{2K\times N_{\mathrm{eff}}}$ and $\mathbf{V}\in\mathbb{C}^{(Q+1)\times N_{\mathrm{eff}}}$
are semi-unitary matrices with $N_{\mathrm{eff}}$ being the rank
of $\mathbf{E}_{\mathrm{oc}}$, also known as effective aerial degrees
of freedom \cite{han2020characteristic,han2022pattern,zhang2025compact} and $\mathbf{S}\in\mathbb{R}^{N_{\mathrm{eff}}\times N_{\mathrm{eff}}}$
collects nonzero singular values of $\mathbf{E}_{\mathrm{oc}}$. Substituting
\eqref{eq:SVD} into \eqref{eq: beamspace channel i}, we can rewrite
the beamspace channel $\mathbf{h}_{u}(\mathbf{b}_{u})$ as 
\begin{equation}
\mathbf{h}_{u}(\mathbf{b}_{u})=\mathbf{w}_{u}^{\mathsf{H}}(\mathbf{b}_{u})\bar{\mathbf{H}}_{u},\forall u,\label{eq:channel}
\end{equation}
where $\mathbf{w}_{u}(\mathbf{b}_{u})\in\mathbb{C}^{N_{\mathrm{eff}}\times1}$
is given by
\begin{equation}
\mathbf{w}_{u}(\mathbf{b}_{u})=\mathbf{S}\mathbf{V}^{\mathsf{T}}\mathbf{i}_{u}^{*}(\mathbf{b}_{u}),\forall u,
\end{equation}
which is referred to as the \textit{pattern coder} for pixel antenna
at user $u$ since it linearly codes the $N_{\mathrm{eff}}$ orthogonal
radiation patterns, i.e. the $N_{\mathrm{eff}}$ columns of $\mathbf{U}$,
to synthesize the radiation pattern of pixel antenna, i.e. $\mathbf{e}_{u}(\mathbf{b}_{u})=\mathbf{U}\mathbf{w}_{u}^{*}(\mathbf{b}_{u})$
and satisfies that $\|\mathbf{w}_{u}(\mathbf{b}_{u})\|_{2}=1$ since
$\|\mathbf{e}_{u}(\mathbf{b}_{u})\|_{2}=1$ and $\mathbf{U}^{\mathsf{H}}\mathbf{U}=\mathbf{I}_{N_\mathrm{eff}}$.
Meanwhile, $\bar{\mathbf{H}}_{u}\in\mathbb{C}^{N_{\mathrm{eff}}\times N}$
is given by
\begin{equation}
\bar{\mathbf{H}}_{u}=\mathbf{U}^{\mathsf{T}}\mathbf{H}_{\mathrm{v},u}\mathbf{E}_{\mathrm{T}},\forall u,
\end{equation}
which can be regarded as an $N_{\mathrm{eff}}\times N$ MIMO channel
matrix. We assume a rich scattering environment with entries $[\mathbf{H}_{\mathrm{v},u}]_{i,j}$
$\forall i,j$ being independent and identically distributed (i.i.d.)
random variables and following $[\mathbf{H}_{\mathrm{v},u}]_{i,j}\sim\mathcal{CN}(0,1)$.
Therefore, leveraging $\mathbf{E}_{\mathrm{T}}^{\mathsf{H}}\mathbf{E}_{\mathrm{T}}=\mathbf{I}_{N}$
and $\mathbf{U}^{\mathsf{H}}\mathbf{U}=\mathbf{I}$, we have that
$[\bar{\mathbf{H}}_{u}]_{i,j}\sim\mathcal{CN}(0,1)$ $\forall i,j$
are also i.i.d. complex Gaussian distributed variables, implying
that $\bar{\mathbf{H}}_{u}$ is essentially the same as the conventional
$N_{\mathrm{eff}}\times N$ MIMO channel with i.i.d. Rayleigh fading. 

With the reformulation in (\ref{eq:channel}), we can better understand
where the benefit of pixel antenna over the conventional antenna comes
from, that is each pixel antenna with $N_{\mathrm{eff}}$ orthogonal
radiation patterns can perform like a conventional $N_{\mathrm{eff}}$-antenna
array to provide extra spatial degrees of freedom. Moreover, with
generally $N_{\mathrm{eff}}\ll K$, the above reformulation (\ref{eq:channel})
is also useful in antenna coding design as we can directly adopt the
channel $\bar{\mathbf{H}}_{u}$ with reduced dimension instead of
using the virtual channel $\mathbf{H}_{\mathrm{v},u}$ with high dimension
to reduce the overhead for channel estimation and the computational
complexity for antenna coding optimization.

\section{Joint Precoding and Antenna Coding Design}

\label{sec:joint_opt}

In this section, we formulate the joint optimization of the precoding
at the transmitter and the antenna coding at the users to maximize
the sum rate of the MU-MISO system using pixel antennas and propose
an alternating precoding and antenna coding optimization algorithm
as follows. 

\subsection{Problem Formulation}

Based on the reformulated beamspace channel \eqref{eq:channel}, the
received signal at user $u$ is given by 
\begin{equation}
y_{u}=\mathbf{w}_{u}^{\mathsf{H}}(\mathbf{b}_{u})\bar{\mathbf{H}}_{u}\sum_{i\in\mathcal{U}}\mathbf{p}_{i}s_{i}+n_{u},\forall u,
\end{equation}
where $n_{u}\sim\mathcal{CN}(0,\sigma_{u}^{2})$, $\forall u\in\mathcal{U}$
denotes the additive white Gaussian noise. Accordingly, the signal-to-interference-plus-noise
(SINR) at user $u$ is defined as 
\begin{equation}
\gamma_{u}(\mathbf{P},\mathbf{b}_{u})=\frac{|\mathbf{w}_{u}^{\mathsf{H}}(\mathbf{b}_{u})\bar{\mathbf{H}}_{u}\mathbf{p}_{u}|^{2}}{\sum_{j\ne u}|\mathbf{w}_{u}^{\mathsf{H}}(\mathbf{b}_{u})\bar{\mathbf{H}}_{u}\mathbf{p}_{j}|^{2}+\sigma_{u}^{2}},\forall u.
\end{equation}
To explore the performance boundary of the considered multi-user MISO
system using pixel antennas, we assume perfect channel state information
is known at the transmitter\footnote{In practice, the channels $\bar{\mathbf{H}}_{u}$ $\forall u$ in
the MU-MISO system can be estimated sequentially. Specifically, the
channel $\bar{\mathbf{H}}_{u}$ for user $u$ can be estimated using
an approach based on beamspace pilot transmission \cite{zhang2022analog}.}. In this case, the sum rate maximization problem is formulated as
\begin{subequations}
    \label{eq:p0_perfectCSIT}
    \begin{align}
        \max_{\mathbf{P},\mathbf{B},\mathbf{i}_{\mathrm{A}}} & R(\mathbf{P},\mathbf{B})\\
        \mathrm{s.t.~~} & \|\mathbf{w}_{u}(\mathbf{b}_{u})\|_{2}=1,\forall u,\label{eq:constraint_w0}\\
        & \mathbf{w}_{u}(\mathbf{b}_{u})=\mathbf{S}\mathbf{V}^{\mathsf{T}}\mathbf{i}_{u}^{*}(\mathbf{b}_{u}),\forall u,\label{eq:constraint_w}\\
        & \mathbf{i}_{u}(\mathbf{b}_{u})=\left[\begin{matrix}1\\
        -(\mathbf{Z}_{\mathrm{PP}}+\mathbf{Z}_{\mathrm{L}}(\mathbf{b}_{u}))^{-1}\mathbf{z}_{\mathrm{PA}}
        \end{matrix}\right]i_{\mathrm{A},u},\forall u,\label{eq:structure_i}\\
        & \mathbf{b}_{u}\in\{0,1\}^{Q\times1},\forall u,\label{eq:constraint_coder}\\
        & \|\mathbf{P}\|_{\mathsf{F}}^{2}\le P, \label{eq:constraint_p}
    \end{align}
\end{subequations}
where $R(\mathbf{P},\mathbf{B})=\sum_{u\in\mathcal{U}}\log_{2}(1+\gamma_{u}(\mathbf{P},\mathbf{b}_{u}))$
with $\mathbf{B}=[\mathbf{b}_{1},\ldots,\mathbf{b}_{U}]$ and $\mathbf{i}_{\mathrm{A}}=[i_{\mathrm{A},1},\ldots,i_{\mathrm{A},U}]^{\mathsf{T}}$
with $i_{\mathrm{A},u}$ being the current at each antenna port for
user $u$ and an optimization variable to ensure the radiation pattern
is normalized.

\subsection{Optimization Algorithm}

Problem (\ref{eq:p0_perfectCSIT}) is a multi-variable optimization
with a sum-of-function-of-ratio objective function, where the variables
are highly coupled. To efficiently solve this problem, we adopt
fractional programming theory \cite{shen2018fractional} and rewrite
the objective function as $R(\mathbf{P},\mathbf{B})=\max_{\bm{\iota},\bm{\tau}}\bar{R}(\mathbf{P},\mathbf{B},\bm{\iota},\bm{\tau})$,
where we have $\bm{\iota}=[\iota_{1},\ldots,\iota_{U}]^{\mathsf{T}}$,
$\bm{\tau}=[\tau_{1},\ldots,\tau_{U}]^{\mathsf{T}}$, and 
\begin{equation}
    \begin{aligned}
        \tilde{R}(\mathbf{P}, & \mathbf{B},\bm{\iota},\bm{\tau})=\sum_{u\in\mathcal{U}}\Big(\log_{2}(1+\iota_{u})-\iota_{u}\\
        & +2\sqrt{1+\iota_{u}}\Re\{\tau_{u}^{*}\mathbf{w}_{u}^{\mathsf{H}}(\mathbf{b}_{u})\bar{\mathbf{H}}_{u}\mathbf{p}_{u}\}\\
         &-|\tau_{u}|^{2}\big(\sum_{p\in\mathcal{U}}|\mathbf{w}_{u}^{\mathsf{H}}(\mathbf{b}_{u})\bar{\mathbf{H}}_{u}\mathbf{p}_{p}|^{2}+\sigma_{u}^{2}\big)\Big).
    \end{aligned}
\end{equation}
As such, problem (\ref{eq:p0_perfectCSIT}) can be rewritten as 
\begin{equation}
\begin{aligned} & \{\mathbf{P}^{\star},\mathbf{B}^{\star},\mathbf{i}_{\mathrm{A}}^{\star},\bm{\iota}^{\star},\bm{\tau}^{\star}\}=\mathop{\mathrm{arg~max}}_{\text{(\ref{eq:constraint_w0})-(\ref{eq:constraint_p})}}\tilde{R}(\mathbf{P},\mathbf{B},\bm{\iota},\bm{\tau}).\end{aligned}
\label{eq:p1_perfectCSIT}
\end{equation}
We propose to iteratively update each variable block in (\ref{eq:p1_perfectCSIT})
with fixed other blocks until convergence. The update of each variable
block is detailed below.

\subsubsection{Update of $\bm{\iota}$ and $\bm{\tau}$}

The $\bm{\iota}$-subproblem and $\bm{\tau}$-subproblem are both
unconstrained convex optimization, yielding the following closed-form
solutions: 
\begin{equation}
\iota_{u}^{\star}=\gamma_{u}(\mathbf{P},\mathbf{b}_{u}),\forall u,
\end{equation}
\begin{equation}
\tau_{u}^{\star}=\frac{\sqrt{1+\iota_{u}}\mathbf{w}_{u}^{\mathsf{H}}(\mathbf{b}_{u})\bar{\mathbf{H}}_{u}\mathbf{p}_{u}}{\sum_{p\in\mathcal{U}}|\mathbf{w}_{u}^{\mathsf{H}}(\mathbf{b}_{u})\bar{\mathbf{H}}_{u}\mathbf{p}_{p}|^{2}+\sigma_{u}^{2}},\forall u.
\end{equation}

\subsubsection{Update of $\mathbf{P}$}

The $\mathbf{P}$-subproblem is given by 
\begin{equation}
\mathbf{P}^{\star}=\arg\max_{\|\mathbf{P}\|_{\mathsf{F}}^{2}\le P}\sum_{u\in\mathcal{U}}(2\Re\{\mathbf{a}_{u}^{\mathsf{H}}\mathbf{p}_{u}\}-\mathbf{p}_{u}^{\mathsf{H}}\mathbf{A}\mathbf{p}_{u}),
\end{equation}
where we have 
\begin{subequations}
\begin{align}\mathbf{A} & =\sum_{p\in\mathcal{U}}\bar{\mathbf{H}}_{p}^{\mathsf{H}}\mathbf{w}_{u}(\mathbf{b}_{p})\mathbf{w}_{u}^{\mathsf{H}}(\mathbf{b}_{p})\bar{\mathbf{H}}_{p}|\tau_{p}|^{2},\\
\mathbf{a}_{u} & =\sqrt{1+\iota_{u}}\bar{\mathbf{H}}_{u}^{\mathsf{H}}\mathbf{w}_{u}(\mathbf{b}_{u})\tau_{u},\forall u.
\end{align}
\end{subequations}
Applying the Lagrangian multiplier method, the closed-form solution
can be given as 
\begin{equation}
\mathbf{p}_{u}^{\star}=(\mathbf{A}+\mu^{\star}\mathbf{I}_{N})^{-1}\mathbf{a}_{u},\forall u,
\end{equation}
where $\mu^{\star}$ is obtained by a bisection search.

\subsubsection{Update of $\mathbf{B}$ and $\mathbf{i}_{\mathrm{A}}$}

The $\{\mathbf{B},\mathbf{i}_{\mathrm{A}}\}$-subproblem is separable
between users and each has the form 
\begin{equation}
    \begin{aligned}
        \max_{\mathbf{b}_{u},i_{\mathrm{A},u}}~ & 2\Re\{\mathbf{w}_{u}^{\mathsf{H}}(\mathbf{b}_{u})\mathbf{q}_{u}\}-\mathbf{w}_{u}^{\mathsf{H}}(\mathbf{b}_{u})\mathbf{Q}_{u}\mathbf{w}_{u}(\mathbf{b}_{u})\\
        \mathrm{s.t.}~~ & \text{(\ref{eq:constraint_w0})-(\ref{eq:constraint_coder})},
    \end{aligned}
\label{eq:subp_b}
\end{equation}
where we have 
\begin{subequations}
\begin{align}
    \mathbf{Q}_{u} & =|\tau_{u}|^{2}\bar{\mathbf{H}}_{u}\sum_{p\in\mathcal{U}}\mathbf{p}_{p}\mathbf{p}_{p}^{\mathsf{H}}\bar{\mathbf{H}}_{u}^{\mathsf{H}},\forall u,\\
\mathbf{q}_{u} & =\sqrt{1+\iota_{u}}\tau_{u}^{*}\bar{\mathbf{H}}_{u}\mathbf{p}_{u},\forall u.
\end{align}
\end{subequations}
From constraints (\ref{eq:constraint_w0})-(\ref{eq:structure_i}),
we observe that, once the antenna coder is determined, $i_{\mathrm{A},u}$
$\forall u\in\mathcal{U}$ can be simply obtained by the normalization
in (\ref{eq:constraint_w0}). In this sense, we equivalently transform
problem (\ref{eq:subp_b}) into the following form 
\begin{subequations}
\label{eq:subp_b1} 
\begin{align}
\max_{\mathbf{b}_{u}}~ & 2\Re\{\mathbf{w}_{u}^{\mathsf{H}}(\mathbf{b}_{u})\mathbf{q}_{u}\}-\mathbf{w}_{u}^{\mathsf{H}}(\mathbf{b}_{u})\mathbf{Q}_{u}\mathbf{w}_{u}(\mathbf{b}_{u})\\
\mathrm{s.t.}~ & \mathbf{w}_{u}(\mathbf{b}_{u})=\frac{\mathbf{S}\mathbf{V}^{\mathsf{T}}\bar{\mathbf{i}}_{u}^{*}(\mathbf{b}_{u})}{\|\mathbf{S}\mathbf{V}^{\mathsf{T}}\bar{\mathbf{i}}_{u}^{*}(\mathbf{b}_{u})\|_{2}},\forall u,\label{eq:w_new}\\
 & \bar{\mathbf{i}}_{u}(\mathbf{b}_{u})=\left[\begin{matrix}1\\
-(\mathbf{Z}_{\mathrm{PP}}+\mathbf{Z}_{\mathrm{L}}(\mathbf{b}_{u}))^{-1}\mathbf{z}_{\mathrm{PA}}
\end{matrix}\right],\forall u,\label{eq:i_bar}\\
 & \mathbf{b}_{u}\in\{0,1\}^{Q\times1},\forall u,\label{eq:coder1}
\end{align}
\end{subequations}
 which is essentially a binary optimization with $\mathbf{b}_{u}$
$\forall u\in\mathcal{U}$ being the only optimization variable.

Problem (\ref{eq:subp_b1}) can be solved by some searching methods,
such as the SEBO proposed
in \cite{shen2016successive}. Taking the design of $\mathbf{b}_{u}$
in (\ref{eq:subp_b1}) as an example, the main steps of SEBO is summarized
as follows.

\begin{enumerate}[{Step 1}.1:]
\item Split the binary variables in the antenna coder $\mathbf{b}_{u}$
into $\lceil\frac{Q}{J}\rceil$ blocks with each block having $J$
binary variables. Cyclically optimize each block by exhaustive search
until convergence of the objective function is guaranteed. 
\item With the optimized antenna coder in Step 1.1, randomly flip up to $J$
bits to check if there is any other local optimum resulting a larger
objective value. 
\end{enumerate}
Therefore, the SEBO algorithm requires $\mathcal{O}(I_{\mathrm{e}}2^{J})$
computational complexity, where $I_{\mathrm{e}}$ denotes the number
of iterations to perform exhaustive search.

\textit{Complexity Analysis: } The complexity of the precoder design
mainly comes from the matrix inverse, which requires complexity $\mathcal{O}(UI_{\mathrm{b}}N^{3})$
with $I_{\mathrm{b}}$ being the number of iterations for the bisection
search. Meanwhile, the complexity of the antenna coder design mainly
comes from the SEBO and is given by $\mathcal{O}(UI_{\mathrm{e}}2^{J})$.
Therefore, the complexity of the proposed alternating algorithm is
$\mathcal{O}(IU(I_{\mathrm{b}}N^{3}+I_{\mathrm{e}}2^{J}))$, where
$I$ denotes the required number of iterations to guarantee the convergence.

\section{Codebook-Based Antenna Coding Design}

\label{sec:codebook}

The proposed alternating algorithm in Section \ref{sec:joint_opt}
may suffer from high computational complexity due to the iteration
and block-exhaustive search using SEBO. To reduce the computational
complexity, we propose a codebook-based antenna coding design, as
explained in details below.

\subsection{Optimization Framework}

\label{subsec:heuristic_codebook}

Given a codebook for antenna coding shared by all users, defined as
\begin{equation}
\mathcal{C}=\{\mathbf{c}_{m}\in\{0,1\}^{Q\times1}~|~\forall m\in\mathcal{M}=\{1,\ldots,M\}\},
\end{equation}
where $M=2^{D}$ with $D$ denoting the number of quantization bits,
we formulate the following problem\footnote{From here we omit the optimization of currents $i_{\mathrm{A},u}$
$\forall u$ since they can be simply obtained by normalization with
given antenna coders as have been explained in (\ref{eq:subp_b1}).} 
\begin{equation}
\begin{aligned}\max_{\mathbf{P},\mathbf{B}}~~ & R(\mathbf{P},\mathbf{B})\\
\mathrm{s.t.~~~} & \text{(\ref{eq:w_new}),~(\ref{eq:i_bar}),}\\
 & \mathbf{b}_{u}\in\mathcal{C},\forall u,\\
 & \|\mathbf{p}_{u}\|_{2}^{2}=\frac{P}{U},\forall u,
\end{aligned}
\label{eq:p1}
\end{equation}
where we assume a uniform power allocation for precoder design for
simplicity. To efficiently solve problem (\ref{eq:p1}), we propose
a heuristic algorithm, whose main idea is to successively update the
zero-forcing (ZF) precoder and each antenna coder by performing a
one-dimensional exhaustive search over the codebook $\mathcal{C}$.
The detailed procedure is summarized as the following steps.

\begin{enumerate}[{Step 2}.1:]
    \item Initialize antenna coders $\mathbf{b}_{1}^{\star},\ldots,\mathbf{b}_{U}^{\star}$
    by randomly choosing from the codebook $\mathcal{C}$, calculate the
    effective channel matrix for all users as $\mathbf{H}_{\mathrm{eff}}^{\star}=[\mathbf{h}_{\mathrm{eff},1}^{\star\mathsf{T}},\ldots,\mathbf{h}_{\mathrm{eff},U}^{\star\mathsf{T}}]^{\mathsf{T}}$,
    where $\mathbf{h}_{\mathrm{eff},u}^{\star}=\mathbf{w}_{u}^{\mathsf{H}}(\mathbf{b}_{u}^{\star})\bar{\mathbf{H}}_{u}$
    $\forall u\in\mathcal{U}$, and obtain the corresponding ZF precoder
    $\mathbf{P}$ with uniform power allocation, i.e. $\mathbf{P}=\sqrt{\frac{P}{U}}[\frac{\bar{\mathbf{p}}_{1}}{\|\bar{\mathbf{p}}_{1}\|_{2}},\ldots,\frac{\bar{\mathbf{p}}_{U}}{\|\bar{\mathbf{p}}_{U}\|_{2}}]$,
    where $\bar{\mathbf{p}}_{u}=[\mathbf{H}_{\mathrm{eff}}^{\star\mathsf{H}}(\mathbf{H}_{\mathrm{eff}}^{\star}\mathbf{H}_{\mathrm{eff}}^{\star\mathsf{H}})^{-1}]_{:,u}$
    $\forall u\in\mathcal{U}$.
    \item For user $u$, calculate the effective channels set 
    \begin{equation}
    \mathcal{H}_{\mathrm{eff},u}=\{\mathbf{h}_{\mathrm{eff},u}^{m}=\mathbf{w}_{u}^{\mathsf{H}}(\mathbf{c}_{m})\bar{\mathbf{H}}_{u}~|~\forall m\in\mathcal{M}\},
    \end{equation}
    using the codewords from $\mathcal{C}$, each of which is an antenna coder with a specific configuration. Obtain the 
    ZF precoders $\mathbf{P}_{u}^{m}=[\mathbf{p}_{u,1}^{m},\ldots,\mathbf{p}_{u,U}^{m}]=\sqrt{\frac{P}{U}}[\frac{\bar{\mathbf{p}}_{u,1}^{m}}{\|\bar{\mathbf{p}}_{u,1}^{m}\|_{2}},\ldots,\frac{\bar{\mathbf{p}}_{u,U}^{m}}{\|\bar{\mathbf{p}}_{u,U}^{m}\|_{2}}]$
    $\forall m\in\mathcal{M}$ with uniform power allocation, where $\bar{\mathbf{p}}_{u,i}^{m}=[\mathbf{H}_{\mathrm{eff},u}^{m\mathsf{H}}(\mathbf{H}_{\mathrm{eff},u}^{m}\mathbf{H}_{\mathrm{eff},u}^{m\mathsf{H}})^{-1}]_{:,i}$
    $\forall i\in\mathcal{U}$ with $\mathbf{H}_{\mathrm{eff},u}^{m}=[\mathbf{h}_{\mathrm{eff},1}^{\star\mathsf{T}},\ldots,\mathbf{h}_{\mathrm{eff},u-1}^{\star\mathsf{T}},\mathbf{h}_{\mathrm{eff},u}^{m\mathsf{T}},\mathbf{h}_{\mathrm{eff},u+1}^{\star\mathsf{T}},\ldots,\mathbf{h}_{\mathrm{eff},U}^{\star\mathsf{T}}]^{\mathsf{T}}$
    being the effective channel matrix for the $U$ users and including 
    \begin{enumerate}
        \item those have been updated in this iteration (if any), i.e. $\mathbf{h}_{\mathrm{eff},1}^{\star},\ldots,\mathbf{h}_{\mathrm{eff},u-1}^{\star}$; 
        \item the effective channel for user $u$ to be determined, i.e. $\mathbf{h}_{\mathrm{eff},u}^{m}\in\mathcal{H}_{\mathrm{eff},u}$; 
        \item and those have been updated in the last iteration (if any), i.e. $\mathbf{h}_{\mathrm{eff},u+1}^{\star},\ldots,\mathbf{h}_{\mathrm{eff},U}^{\star}$.
    \end{enumerate}
    \item Update $\mathbf{b}_{u}\in\mathcal{C}$ and $\mathbf{h}_{\mathrm{eff},u}\in\mathcal{H}_{\mathrm{eff},u}$
    by finding the index of the codeword leading to the maximum sum rate,
    i.e. 
    \begin{equation}
        \begin{aligned}
            m_{u}^{\star}=\arg & \max_{m\in\mathcal{M}}~\log_{2}\left(1+\frac{|\mathbf{h}_{\mathrm{eff},u}^{m\mathsf{H}}\mathbf{p}_{u,u}^{m}|^{2}}{\sigma_{u}^{2}}\right)\\
            &+\sum_{j\ne u}\log_{2}\left(1+\frac{|\mathbf{h}_{\mathrm{eff},j}^{\star\mathsf{H}}\mathbf{p}_{u,j}^{m}|^{2}}{\sigma_{j}^{2}}\right).
        \end{aligned}
    \end{equation}
    This is done by a one-dimensional exhaustive search, which yields
    the antenna coder $\mathbf{b}_{u}^{\star}=\mathbf{c}_{m_{u}^{\star}}$
    and the corresponding effective channel $\mathbf{h}_{\mathrm{eff},u}^{\star}=\mathbf{w}_{u}^{\mathsf{H}}(\mathbf{b}_{u}^{\star})\bar{\mathbf{H}}_{u}$. 
    \item Repeat Steps 2.2 and 2.3 for all users $u\in\mathcal{U}$. 
    \item Repeat Step 2.4 for all iterations until the convergence of the sum rate
    is achieved. 
\end{enumerate}
Compared to the alternating design proposed in Section \ref{sec:joint_opt}
where the update of each antenna coder in each iteration requires
complexity $\mathcal{O}(I_{\mathrm{e}}2^{J})$, the update of each
antenna coder in the heuristic codebook-based design only needs complexity
$\mathcal{O}(2^{D})$. Given that the SEBO generally requires many
iterations and a large block size to update all the bits and guarantee
a satisfactory performance, the heuristic codebook-based design algorithm
can be much more efficient than the alternating algorithm in Section
\ref{sec:joint_opt}.

\textit{Complexity Analysis:} The complexity of the above algorithm
mainly comes from how many times the matrix inverse is calculated
in each iteration to obtain the ZF precoder, i.e. $U2^{M}$ times.
Therefore, the complexity of the whole procedure is given by $\mathcal{O}(I'2^{D}U^{4})$, where $I'$ denotes the number of iterations to guarantee convergence.

\subsection{Codebook Design}

\label{subsec:codebook_design}

The key of the proposed heuristic algorithm in Section \ref{subsec:heuristic_codebook}
is to design an efficient codebook $\mathcal{C}$, the solution to
which will be detailed below.

To facilitate the codebook design, we assume the channels from the
transmitter to all users follow the same distribution, and generate
a common training set
\begin{equation}
    \begin{aligned}
        \mathcal{H} & =\{\bar{\mathbf{H}}^{s}=[\bar{\mathbf{H}}_{1}^{s},\ldots,\bar{\mathbf{H}}_{U}^{s}]\in\mathbb{C}^{N_{\mathrm{eff}}\times NU}~|~\forall s=1,\ldots,S\},
    \end{aligned}\label{eq:set}
\end{equation}
where $S$ denotes the number of channel realizations in the training
set. Given a channel realization $\bar{\mathbf{H}}^{s}$ and the selected
codeword $\mathbf{c}_{m}$ (shared for all users), we define the following
sum rate performance metric
\begin{equation}
    \bar{R}(\bar{\mathbf{H}}^{s},\mathbf{c}_{m})=\sum_{\forall u\in\mathcal{U}}\log_{2}(1+\bar{\rho}|\mathbf{w}_u^{\mathsf{H}}(\mathbf{c}_{m})\bar{\mathbf{H}}_{u}^{s}\bar{\mathbf{p}}_{u}(\bar{\mathbf{H}}^{s},\mathbf{c}_{m})|^{2}),
\end{equation}
where $\bar{\rho}$ represents the SNR, while
it is independent of the transmit power $P$ for the aim to generate
a codebook commonly used for different power budgets. $\bar{\mathbf{p}}_{u}(\bar{\mathbf{H}}^{s},\mathbf{c}_{m})$
$\forall u\in\mathcal{U}$ refer to the normalized ZF precoder vectors
which perfectly cancel the inter-user interference. Therefore, each
precoder is a function of the channel realization $\bar{\mathbf{H}}^{s}$
and the codeword $\mathbf{c}_{m}$. Given a codebook $\mathcal{C}$,
selecting the best codewords that maximize the performance metric
$\bar{R}(\bar{\mathbf{H}}^{s},\mathbf{c}_{m})$ for all $\bar{\mathbf{H}}^{s}\in\mathcal{H}$
essentially partitions $\mathcal{H}$ into $M$ subsets $\mathcal{H}_{1},\ldots,\mathcal{H}_{M}$,
with $\mathcal{H}_{m}$ representing the nearest neighbor of the codeword
$\mathbf{c}_{m}$ given by 
\begin{equation}
    \mathcal{H}_{m}=\{\bar{\mathbf{H}}^{s}|\bar{R}(\bar{\mathbf{H}}^{s},\mathbf{c}_{m})\ge\bar{R}(\bar{\mathbf{H}}^{s},\mathbf{c}_{m'}),\forall m\ne m',\forall s\}.\label{eq:partition}
\end{equation}
Then we can formulate the following codebook optimization problem
as 
\begin{subequations}
    \label{eq:p2_codebook} 
    \begin{align}
        \max_{\mathcal{C}}~ & \frac{1}{S}\sum_{m\in\mathcal{M}}\sum_{\bar{\mathbf{H}}^{s}\in\mathcal{H}_{m}}\bar{R}(\bar{\mathbf{H}}^{s},\mathbf{c}_{m})\\
        \mathrm{s.t.}~ & \mathbf{w}_u(\mathbf{c}_{m})=\frac{\mathbf{S}\mathbf{V}^{\mathsf{T}}\bar{\mathbf{i}}_u^{*}(\mathbf{c}_{m})}{\|\mathbf{S}\mathbf{V}^{\mathsf{T}}\bar{\mathbf{i}}_u^{*}(\mathbf{c}_{m})\|_{2}},\forall m,\label{eq:constraint_wc}\\
        & \bar{\mathbf{i}}_u(\mathbf{c}_{m})=\left[\begin{matrix}1\\
        -(\mathbf{Z}_{\mathrm{PP}}+\mathbf{Z}_{\mathrm{L}}(\mathbf{c}_{m}))^{-1}\mathbf{z}_{\mathrm{PA}}
        \end{matrix}\right],\forall m,\label{eq:structure_ic}\\
        & \mathbf{c}_{m}\in\{0,1\}^{Q\times1},\forall m.
    \end{align}
\end{subequations}

Problem (\ref{eq:p2_codebook}) can be viewed as a vector quantization
problem, which can be solved by the generalized Lloyd algorithm \cite{xia2006design}
with the main idea of alternatively optimizing the partition of the
training set and codewords. The procedure is briefly summarized as
follows.

\begin{enumerate}[Step 3.1:]
    \item Properly initialize the codebook $\mathcal{C}$. 
    \item \textit{Nearest Neighbor Partitioning.} Associated with determined
    codewords, partition the training set into $M$ subsets following
    the rule in (\ref{eq:partition}). 
    \item \textit{Centroid Calculation.} The centroid of each partition $\mathcal{H}_{m}$
    is designed to maximize the average performance for this partition.
    Hence, the objective of this step is to calculate the new codewords
    $\mathbf{c}_{m}$, $\forall m\in\mathcal{M}$, each of which independently
    solves 
    \begin{equation}
        \begin{aligned}
            \max_{\mathbf{c}_{m}}~ & \frac{1}{|\mathcal{H}_{m}|}\sum_{\bar{\mathbf{H}}^{s}\in\mathcal{H}_{m}}\bar{R}(\bar{\mathbf{H}}^{s},\mathbf{c}_{m})\\
            \mathrm{s.t.}~ & \text{(\ref{eq:constraint_wc}),~(\ref{eq:structure_ic}),~}\mathbf{c}_{m}\in\{0,1\}^{Q\times1},
        \end{aligned}
        \label{eq:p_codeword}
    \end{equation}
    where the objective function has the only variable $\mathbf{c}_{m}$.
    Therefore, this binary optimization can be directly solved by the
    SEBO algorithm. 
    \item Repeat Steps 3.2 and 3.3 until the convergence of the objective in
    (\ref{eq:p2_codebook}) is achieved. 
\end{enumerate}
Following the above steps, we can obtain the codebook $\mathcal{C}^{\star}=\{\mathbf{c}_{1}^{\star},\ldots,\mathbf{c}_{M}^{\star}\}$
to facilitate the proposed heuristic codebook-based algorithm in Section
\ref{subsec:heuristic_codebook}.

\section{Hierarchical Codebook Design for \\ Antenna Coding}

\label{sec:hierarchical_codebook}

The performance of the codebook-based algorithm proposed in Section
\ref{sec:codebook} is highly related to the resolution of the codebook.
That is, a larger $M$ has a higher probability to increase the sum rate
performance. However, this is achieved at the cost of an exhaustive
search among numerous codewords, which will increase the computational
complexity. To further reduce the computational complexity, in this
section, we propose a hierarchical codebook design for antenna coding.
Below we will first elaborate on how the hierarchical codebook is
used to design the antenna coding, and then illustrate how to design
the hierarchical codebook.

\subsection{Structure of the Hierarchical Codebook}

\begin{figure}
\centering \includegraphics[width=0.48\textwidth]{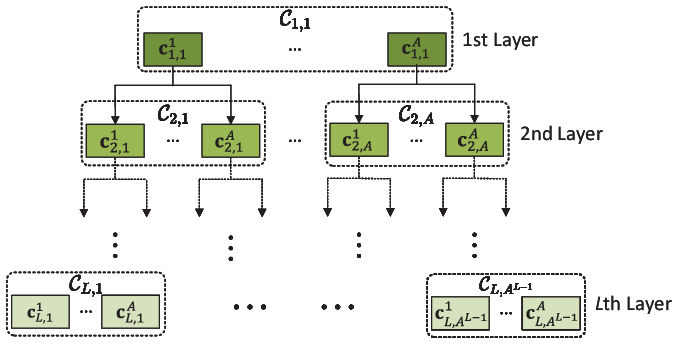}
\caption{A hierarchical codebook $\mathcal{C}$ with $L$ layers.}
\label{fig:hierarchical_codebook} 
\end{figure}

A hierarchical codebook for antenna coding shared by all users, $\mathcal{C}$,
is structured as a tree consisting of $L$ layers, as illustrated
in Fig. \ref{fig:hierarchical_codebook}. Specifically, the $l$th
layer, $\forall l=1,\ldots,L$, consists of $A^{l-1}$ sub-codebooks,
$\mathcal{C}_{l,1},\ldots,\mathcal{C}_{l,A^{l-1}}$, each of which
contains $A$ codewords, i.e. 
\begin{equation}
\mathcal{C}_{l,i}=\{\mathbf{c}_{l,i}^{a}\in\{0,1\}^{Q\times1}~|~\forall a=1,\ldots,A\},
\end{equation}
$\forall i=1,\ldots,A^{l-1}$. Therefore, the hierarchical codebook
is constructed as 
\begin{equation}
\mathcal{C}=\bigcup_{l=1}^{L}\mathcal{C}_{l},~\mathcal{C}_{l}=\bigcup_{i=1}^{A^{l-1}}\mathcal{C}_{l,i}.
\end{equation}
In the $L$-layer hierarchical codebook, adjacent layers are linked
to each other similar to a tree. That is, each codeword in the $l$th
layer is linked to a descendent sub-codebook in the $(l+1)$th layer,
i.e. 
\begin{equation}
\mathbf{c}_{l,i}^{a}\rightarrow\mathcal{C}_{l+1,A(i-1)+a},\forall a=1,\ldots,A.\label{eq:link}
\end{equation}
For example, given $A=3$, $l=2$, $i=3$, and $a=1$, the codeword
$\mathbf{c}_{2,3}^{1}$ is linked to its descendent sub-codebook $\mathcal{C}_{3,7}$,
i.e. $\mathbf{c}_{2,3}^{1}\rightarrow\mathcal{C}_{3,7}$.

\subsection{Optimization Framework Using the Hierarchical Codebook}

\label{subsec:heuristic_hierarchical_codebook}

Based on the hierarchical codebook, we propose a heuristic algorithm
to jointly optimize the precoding and antenna coding, with the main
idea of successively updating the ZF precoder and each antenna coder
by performing a hierarchical binary search over the hierarchical codebook
$\mathcal{C}$. The detailed procedure is summarized below.

\begin{enumerate}[Step 4.1:]
    \item Initialize antenna coders $\mathbf{b}_{1}^{\star},\ldots,\mathbf{b}_{U}^{\star}$ by randomly choosing from the sub-codebook $\mathcal{C}_{1}$, calculate the effective channels $\mathbf{h}_{\mathrm{eff},1}^{\star},\ldots,\mathbf{h}_{\mathrm{eff},U}^{\star}$, and obtain the corresponding ZF precoder $\mathbf{P}$ with uniform power allocation. Initialize $l=1$ and $i=1$. 
    \item For user $u$, in the $l$th layer, calculate effective channels $\mathbf{h}_{\mathrm{eff},u,l,i}^{a}=\mathbf{w}_u^{\mathsf{H}}(\mathbf{c}_{l,i}^{a})\bar{\mathbf{H}}_{u}$,
    $\forall i=1,\ldots,A^{l-1}$, $\forall a=1,\ldots,A$, based on the
    sub-codebook $\mathcal{C}_{l,i}$. Obtain the corresponding ZF precoders
    $\mathbf{P}_{u,l,i}^{a}=[\mathbf{p}_{u,l,i,1}^{a},\ldots,\mathbf{p}_{u,l,i,U}^{a}]$ with uniform power allocation based on the $U$ effective channels including 
    \begin{enumerate}
        \item those have been updated in this iteration (if any), i.e. $\mathbf{h}_{\mathrm{eff},1}^{\star},\ldots,\mathbf{h}_{\mathrm{eff},u-1}^{\star}$; 
        \item the effective channel to be determined, i.e. $\mathbf{h}_{\mathrm{eff},u,l,i}^a\in\{\mathbf{h}_{\mathrm{eff},u,l,i}^{1},\ldots,\mathbf{h}_{\mathrm{eff},u,l,i}^{A}\}$; 
        \item and those have been updated in the last iteration (if any), i.e. $\mathbf{h}_{\mathrm{eff},u+1}^{\star},\ldots,\mathbf{h}_{\mathrm{eff},U}^{\star}$. 
    \end{enumerate}
    Calculate corresponding objective values as 
    \begin{equation}
    \begin{aligned}
        R_{u,l,i}^{a} & =\log_{2}\left(1+\frac{|\mathbf{h}_{\mathrm{eff},u,l,a}^{a\mathsf{H}}\mathbf{p}_{u,l,i,u}^{a}|^{2}}{\sigma_{u}^{2}}\right)\\
        &+\sum_{j\ne u}\log_{2}\left(1+\frac{|\mathbf{h}_{\mathrm{eff},j}^{\star\mathsf{H}}\mathbf{p}_{u,l,i,j}^{a}|^{2}}{\sigma_{j}^{2}}\right),\forall a.
    \end{aligned}
\end{equation}

\item Update the value of $i$ by 
\begin{equation}
i=(i-1)A+\arg\max_{a=1,\ldots,A}R_{u,l,i}^{a},
\end{equation}
and update $l=l+1$. 
\item Repeat Steps 4.2 and 4.3 until $l>L$. Update the antenna coder 
\begin{equation}
\mathbf{b}_{u}^{\star}=\begin{cases}
\mathbf{c}_{L,i}^{\mathsf{mod}(i,A)} & \text{if~}\mathsf{mod}(i,A)\ne0,\\
\mathbf{c}_{L,i}^{A} & \text{else},
\end{cases}
\end{equation}
and the corresponding effective channel $\mathbf{h}_{\mathrm{eff},u}^{\star}=\mathbf{w}_u^{\mathsf{H}}(\mathbf{b}_{u}^{\star})\bar{\mathbf{H}}_{u}$. 
\item Repeat Step 4.4 for all users $u\in\mathcal{U}$. 
\item Repeat Step 4.5 for all iterations until the convergence of the sum rate
is achieved. 
\end{enumerate}
Benefiting from the hierarchical codebook, the update of each antenna
coder needs only $\mathcal{O}(AL)$, which can be much more efficient
than the one-dimensional exhaustive search in the previous section
that requires $\mathcal{O}(2^{D})$.

\textit{Complexity Analysis:} The complexity of the above algorithm
again mainly comes from the time of calculating ZF precoders. Therefore,
the complexity of the whole algorithm is given by $\mathcal{O}(I''ALU^{4})$,
where $I''$ denotes the number of iterations to guarantee convergence.

\subsection{Codebook Design}

The performance of the proposed optimization framework in Section
\ref{subsec:heuristic_hierarchical_codebook} highly depends on an
efficient hierarchical codebook design, while the special structure
and the hierarchical search make sub-codebooks from adjacent layers
link to each other and thus make the design of the whole codebook
different and complicated. Below, we will explain how to design each
sub-codebook based on the solution in \ref{subsec:codebook_design}
and link all sub-codebooks together to construct the hierarchical
codebook $\mathcal{C}$.

\subsubsection{Hierarchical Layer Construction}

The hierarchical search explained in \ref{subsec:heuristic_hierarchical_codebook}
implies the link from each codeword in the $l$th layer to its descendent
subcodebook in the $(l+1)$th layer as illustrated in (\ref{eq:link}).
Moreover, given a common training set $\mathcal{H}$ as defined in (\ref{eq:set}), we first obtain $\mathcal{C}_{1,1}$
based on the sub-codebook design proposed in Section \ref{eq:subcodebook_opt}.
Given a training set $\mathcal{H}_{l,i}\subseteq\mathcal{H}$ and a codebook $\mathcal{C}_{l,i}$, selecting
the best codewords $\mathbf{c}_{l,i}^{a}\in\mathcal{C}_{l,i}$ that
maximize the performance metric $\bar{R}(\bar{\mathbf{H}}^{s},\mathbf{c}_{l,i}^{a})$
for all $\bar{\mathbf{H}}^{s}\in\mathcal{H}_{l,i}$ essentially partitions
$\mathcal{H}_{l,i}$ into $A$ subsets $\mathcal{H}_{l,i}^{1},\ldots,\mathcal{H}_{l,i}^{A}$.
Then, $\mathcal{H}_{l,i}^{a}$ $\forall a$ represents the nearest
neighbor of the codeword $\mathbf{c}_{l,i}^{a}$, that is 
\begin{equation}
    \begin{aligned}
        \mathcal{H}_{l,i}^{a}=\{\bar{\mathbf{H}}^{s}~|~\bar{R}(\bar{\mathbf{H}}^{s},\mathbf{c}_{l,i}^{a}) & \ge\bar{R}(\bar{\mathbf{H}}^{s},\mathbf{c}_{l,i}^{a'}),\\
        &\forall\bar{\mathbf{H}}^{s}\in\mathcal{H}_{l,i},\forall a'\ne a\}.
    \end{aligned}
    \label{eq:partition_binary}
\end{equation}
This motivates us to establish the following link, that is 
\begin{equation}
\mathbf{c}_{l,i}^{a}\rightarrow\mathcal{H}_{l,i}^{a}=\mathcal{H}_{l+1,A(i-1)+a}\rightarrow\mathcal{C}_{l+1,A(i-1)+a}.
\end{equation}
For example, given $A=3$, a codeword $\mathbf{c}_{2,3}^{1}$, and
the common training set $\mathcal{H}_{2,3}$, we can perform (\ref{eq:partition_binary})
to obtain the neighbor set $\mathcal{H}_{2,3}^{1}$, which is used
as the common training set, i.e. $\mathcal{H}_{2,3}^{1}=\mathcal{H}_{3,7}$,
for obtaining the sub-codebook $\mathcal{C}_{3,7}$.

Now we can establish the procedure of constructing the hierarchical
codebook as detailed below.

\begin{enumerate}[Step 5.1:]
\item Initialize $l=1$ and $i=1$. 
\item Obtain the sub-codebook $\mathcal{C}_{1,1}$. 
\item Obtain $\mathcal{H}_{1,1}^{1},\ldots,\mathcal{H}_{1,1}^{A}$ with
$\mathcal{C}_{1,1}$ by (\ref{eq:partition_binary}). 
\item Update partitions 
\begin{equation}
\mathcal{H}_{l+1,A(i-1)+a}=\mathcal{H}_{l,i}^{a},\forall a,
\end{equation}
and obtain the corresponding $A$ sub-codebooks $\mathcal{C}_{l+1,A(i-1)+1},\ldots,\mathcal{C}_{l+1,iA}$. 
\item For each $\mathcal{C}_{l+1,A(i-1)+a}$, $\forall a$, obtain its partitions
$\mathcal{H}_{l+1,A(i-1)+a}^{1},\ldots,\mathcal{H}_{l+1,A(i-1)+a}^{A}$
by (\ref{eq:partition_binary}). 
\item Repeat Steps 5.4 and 5.5 for $i=1,\ldots,A^{l}$. 
\item Update $l=l+1$. 
\item Repeat Steps 5.6 and 5.7 until $l=L$. 
\end{enumerate}

\subsubsection{Sub-codebook $\mathcal{C}_{l,i}$ Design}

\label{eq:subcodebook_opt}

The above procedure decouples the design of the whole hierarchical
codebook into a successive design of the sub-codebook, while the remaining
difficulty is to obtain each sub-codebook. Specifically, when focusing
on the design of sub-codebook $\mathcal{C}_{l,i}$ with a given common
training set $\mathcal{H}_{l,i}$, we can formulate the following
problem based on (\ref{eq:partition_binary}), that is 
\begin{subequations}
    \label{eq:opt_hierarchical_codebook} 
    \begin{align}
        \max_{\mathcal{C}_{l,i}}~ & \frac{1}{|\mathcal{H}_{l,i}|}\sum_{a=1}^{A}\sum_{\bar{\mathbf{H}}^{s}\in\mathcal{H}_{l,i}^{a}}\bar{R}(\bar{\mathbf{H}}^{s},\mathbf{c}_{l,i}^{a})\\
        \mathrm{s.t.}~ & \mathbf{w}_u(\mathbf{c}_{l,i}^{a})=\frac{\mathbf{S}\mathbf{V}^{\mathsf{T}}\bar{\mathbf{i}}_u^{*}(\mathbf{c}_{l,i}^{a})}{\|\mathbf{S}\mathbf{V}^{\mathsf{T}}\bar{\mathbf{i}}_u^{*}(\mathbf{c}_{l,i}^{a})\|_{2}},\forall a,\\
        & \bar{\mathbf{i}}_u(\mathbf{c}_{l,i}^{a})=\left[\begin{matrix}1\\
        -(\mathbf{Z}_{\mathrm{PP}}+\mathbf{Z}_{\mathrm{L}}(\mathbf{c}_{l,i}^{a}))^{-1}\mathbf{z}_{\mathrm{PA}}
        \end{matrix}\right],\forall a,\\
        & \mathbf{c}_{l,i}^{a}\in\{0,1\}^{Q\times1},\forall a.
    \end{align}
\end{subequations}
 Note that problem (\ref{eq:opt_hierarchical_codebook}) has exactly
the same form as (\ref{eq:p2_codebook}). Therefore, we can directly
solve (\ref{eq:opt_hierarchical_codebook}) by the generalized Lloyd
algorithm, i.e. Steps 3.1-3.4, as detailed in Section \ref{subsec:codebook_design}.

\section{Performance Evaluation}

\label{sec:performance}

In this section, we evaluate the performance of the MU-MISO system
using pixel antennas with the joint precoding and antenna coding design.

\subsection{Simulation Setup}

\begin{figure}
\centering \includegraphics[width=0.4\textwidth]{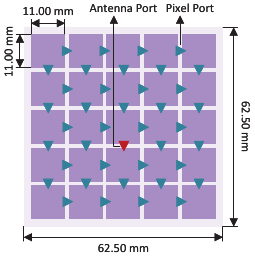}
\caption{Pixel antenna with a physical aperture of $0.5\lambda\times0.5\lambda$
modeled by a multiport network with an antenna port and $Q=39$ pixel
ports.}
\label{fig:demo} 
\end{figure}

We consider a propagation environment with rich scattering to formulate
a 2-D uniform power angular spectrum. The number of sampled spatial
angles is set as $K=72$. At the transmitter, each antenna has a fixed
configuration and an isotropic radiation pattern and the antennas
are spatially separated without mutual coupling and spatial correlations.
The pixel antenna parameter settings are based on \cite{shen2024antenna}.
Specifically, the pixel antenna is operating at 2.4 GHz with wavelength
$\lambda=125$ mm. The pixel antenna with a physical aperture $0.5\lambda\times0.5\lambda$ is designed based on discretizing the radiation surface of a microstrip
patch antenna into a grid of pixels, which leads to a multiport network
model with an antenna port and $Q=39$ pixel ports, as illustrated
in Fig. \ref{fig:demo}. Accordingly, the impedance matrix $\mathbf{Z}\in\mathbb{C}^{(Q+1)\times(Q+1)}$ of the $(Q+1)$-port network and the open-circuit radiation pattern matrix $\mathbf{E}_{\mathrm{oc}}\in\mathbb{C}^{2K\times(Q+1)}$ can
be obtained by simulation using the CST studio suite. The channels
$[\bar{\mathbf{H}}_{u}]_{i,j}$ $\forall i,j$ are assumed to follow
i.i.d. complex Gaussian distribution. The noise powers are set as
$\sigma_{1}^{2}=\ldots=\sigma_{U}^{2}=1$.

\subsection{Radiation Pattern}

\begin{figure}[t]
    \centering 
    \includegraphics[width=0.48\textwidth]{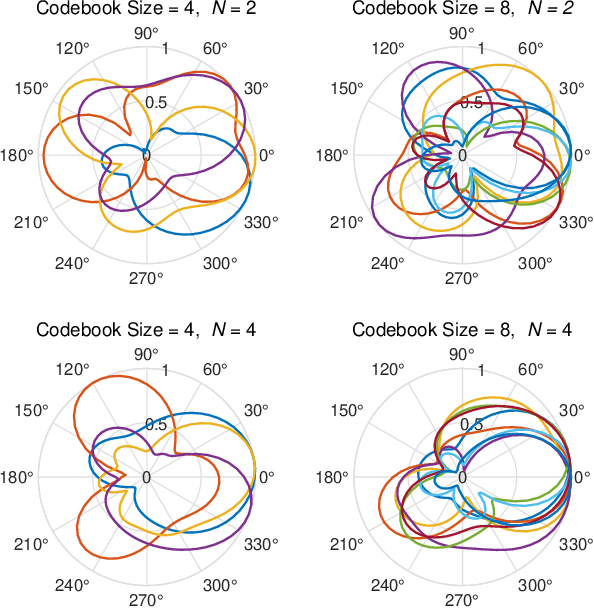}
    \caption{Radiation patterns of the pixel antenna based on various numbers of
    codebook size and transmit antennas.}
    \label{fig:pattern} 
\end{figure}

We start by visualizing the radiation pattern of each pixel antenna
based on the proposed codebooks. Specifically, Fig. \ref{fig:pattern}
illustrates radiation patterns for all codewords from codebooks with
various codebook size constructed based on the proposed codebook design
algorithm in Section \ref{sec:codebook}. It is shown in Fig. \ref{fig:pattern}
that the angle resolution of radiation pattern for the designed codebook
generally increases with the codebook size. This is consistent with
the fact that a larger codebook consists of more diverse and directive
radiation patterns to provide a higher adaptivity to the multiple
paths with different AoAs and AoDs so as to achieve a higher channel
gain while maintaining a lower interference during the data transmission.

\begin{figure}[htbp]
    \centering 
    \includegraphics[width=0.48\textwidth]{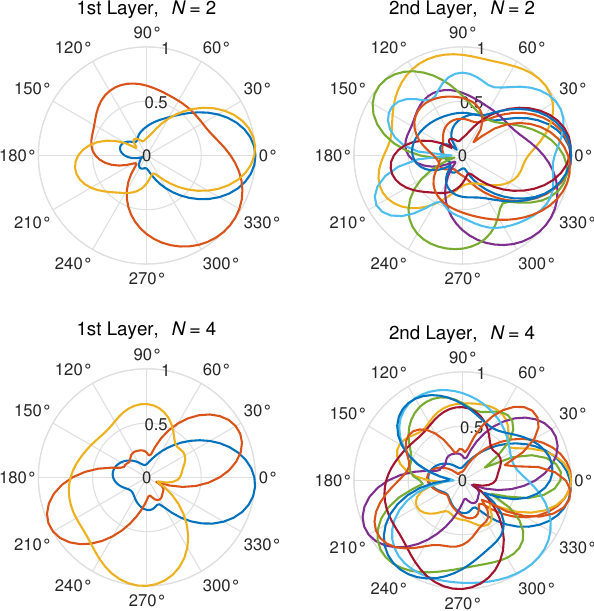}
    \caption{Radiation patterns of the pixel antenna based on a 2-layer hierarchical
    codebook with various numbers of transmit antennas ($A=3$).}
    \label{fig:pattern_hierarchical} 
\end{figure}

In Fig. \ref{fig:pattern_hierarchical}, we further plot the radiation
patterns from a hierarchical codebook with two layers constructed
based on the proposed hierarchical codebook design algorithm in Section
\ref{sec:hierarchical_codebook}. It is shown in Fig. \ref{fig:pattern_hierarchical}
that the radiation patterns for codewords from higher layers tend
to cover wider angle ranges and provide more directivity compared
with the lower layer. This again aligns with the fact that, when constructing
a hierarchical codebook, the codewords in the higher layer are designed
to create a finer division of the beamspace with higher channel gain
and lower interference, so that the antenna coder can be gradually
refined during searching the different levels of hierarchical codebook
and finally the sum rate performance of the MU-MISO system using pixel
antennas can be maximized.

\subsection{Sum Rate Performance}

\begin{figure*}[t]
\centering \subfigure[$N = U = 2$]{\includegraphics[width=0.485\textwidth]{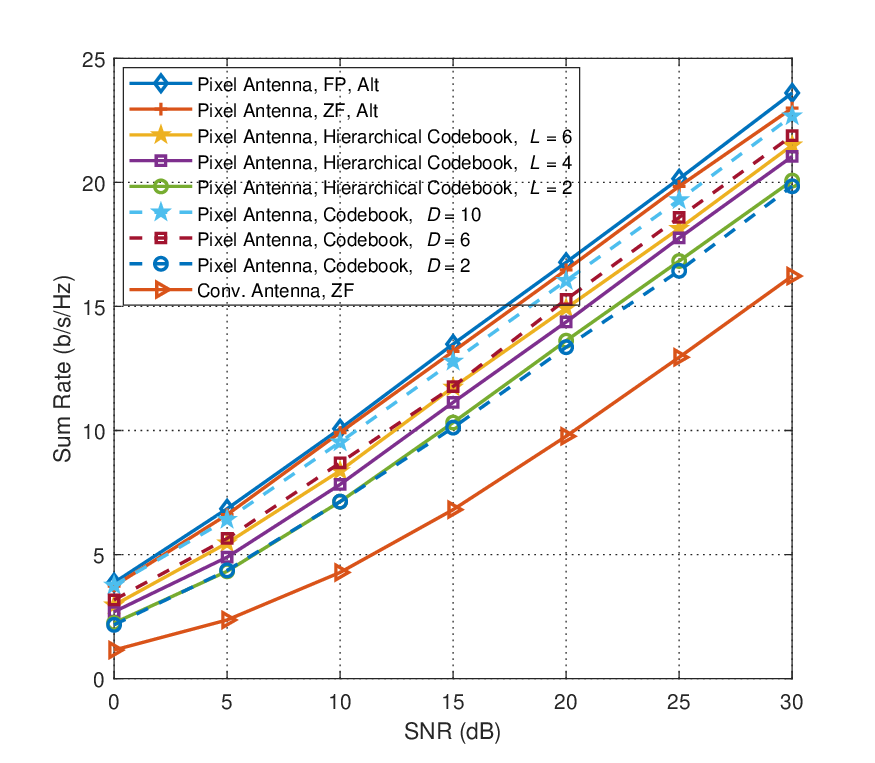}}
\subfigure[$N = U = 4$]{\includegraphics[width=0.485\textwidth]{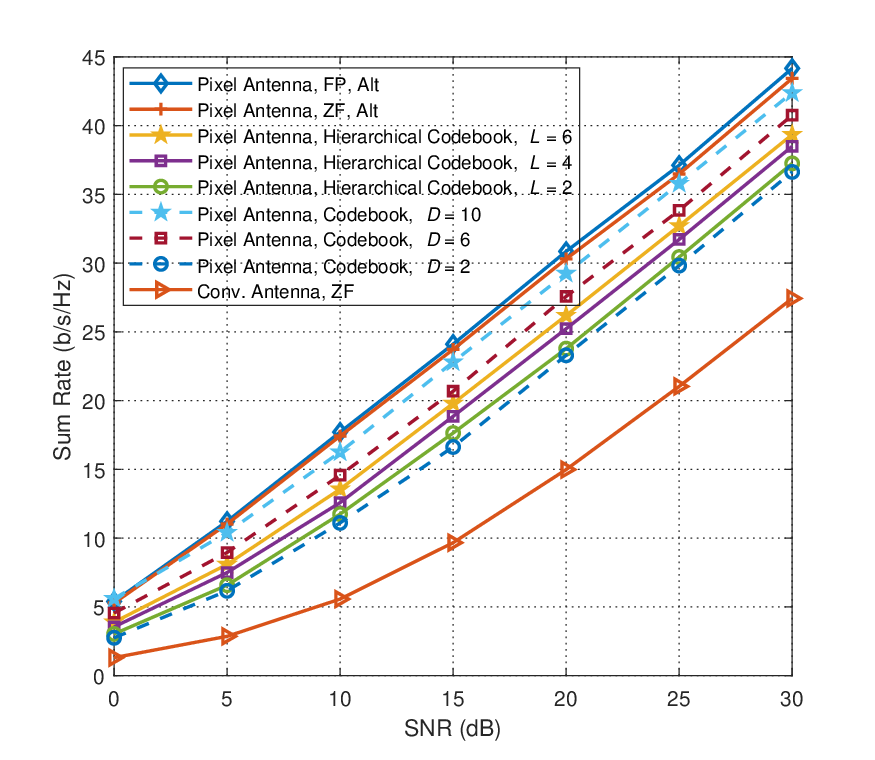}}
\caption{Sum rate versus SNR for MU-MISO systems with conventional antennas
and pixel antennas at users.}
\label{fig:SR_P} 
\end{figure*}

We next evaluate the sum rate performance of the MU-MISO pixel antenna
system. In Fig. \ref{fig:SR_P}, we plot the sum rate performance
as a function of SNR under different parameter settings. For comparison,
we consider different algorithms and benchmarks including
\begin{enumerate}[1)]
    \item \textbf{``FP, Alt''} that refers to the alternating algorithm for precoding design using fractional programming and antenna coding design using
    SEBO proposed in Section \ref{sec:joint_opt};
    \item \textbf{``ZF, Alt''} that refers to the alternating algorithm for precoding design using zero forcing and antenna coding design using SEBO proposed in \cite{spawc2025};
    \item \textbf{``Codebook''} that refers to the joint precoding and codebook-based antenna coding design algorithm proposed in Section \ref{sec:codebook};
    \item \textbf{``Hierarchical Codebook''} that refers to the joint precoding
    and hierarchical codebook-based antenna coding design algorithm proposed
    in Section \ref{sec:hierarchical_codebook};
    \item \textbf{``Conv. Antenna, ZF''} that refers to conventional MU-MISO system
    using conventional antennas at both the transmitter and users where
    the zero-forcing precoding and the water-filling algorithm for power
    allocation are used.
\end{enumerate}
From Fig. \ref{fig:SR_P}, we can make the following observations.

\textit{First}, the proposed MU-MISO system using pixel antennas always
achieves better performance than the conventional MU-MISO system,
thanks to the high reconfigurablity and adaptivity provided by pixel
antennas. For example, when SNR = 10 dB and $N=U=2$, using pixel
antennas at each user (based on the ``FP, Alt'' algorithm) can increase
the sum rate performance by up to 100\% compared to conventional MU-MISO systems using conventional antennas at both sides; when SNR
= 30 dB and $N=U=4$, using pixel antennas at each user can increase
the sum rate performance by up to 67\% compared to conventional MU-MISO systems. These results highlight the advantage of using pixel antennas in significantly
improving the MU-MISO system performance.

\textit{Second}, when optimizing MU-MISO pixel antenna systems, the
``FP, Alt'' algorithm performs better than the ``ZF, Alt'' algorithm
since the ``FP, Alt'' algorithm is based on a rigorous mathematical
optimization with more guaranteed performance, while the latter is
designed heuristically. This also demonstrates that using pixel antennas
does not affect the benefit of fractional programming in optimizing
MU-MISO systems.

\textit{Third}, the performance of the proposed codebook-based algorithm
depends on the codebook size. 
Specifically, with a relatively large codebook size, e.g. with quantization bits
$D=10$, the codebook-based algorithm can achieve satisfactory performance
close to the alternative design algorithm with significantly reduced computational complexity, as will be shown in Fig. \ref{fig:cputime}. 
For example, when SNR =
20 dB and $N=U=4$, the codebook-based algorithm with a codebook size
$D=10$ can achieve around 97\% of the performance achieved by the
``FP, Alt'' algorithm.
Meanwhile, even though with a relatively small
codebook size, the codebook-based algorithm can achieve improved sum rate performance over conventional antenna systems with low computational complexity. 
For example, when SNR = 10 dB and $N=U=2$,
the codebook-based algorithm with a codebook size $D=2$ can still improve the sum rate performance by around 50\% over conventional antenna systems, while requiring only 4 configurations for each antenna coder.
This observation demonstrates the effectiveness
of the proposed codebook-based algorithm with an efficient codebook
design and a properly selected codebook size.

\textit{Fourth}, the performance of the proposed hierarchical codebook-based
algorithm depends on the codebook layers. More importantly, it is
possible to use a hierarchical codebook with limited layers to reach
the performance achieved by a large-dimensional codebook. For example,
when SNR $\le$ 15 dB, the hierarchical codebook-based algorithm with
$L=6$ achieves almost the same performance as the codebook-based
algorithm with $D=6$, while the former (requiring $AL =18$ configurations for each antenna coder) is more computational efficient
than the latter (requiring $2^D = 64$ configurations for each antenna coder).

Overall, using the ``FP, Alt'' algorithm in pixel antenna systems has its unique advantage in achieving always the best performance; using the ``Codebook'' algorithm with a proper codebook size and the ``Hierarchical Codebook'' algorithm with a proper codebook layer in pixel antenna systems have benefits in achieving satisfactory performance closed to that using the ``FP, Alt'' algorithm and over the conventional antenna system.

\subsection{Computational Time}

\begin{figure*}[t]
\centering \subfigure[$N = U = 2$]{\includegraphics[width=0.485\textwidth]{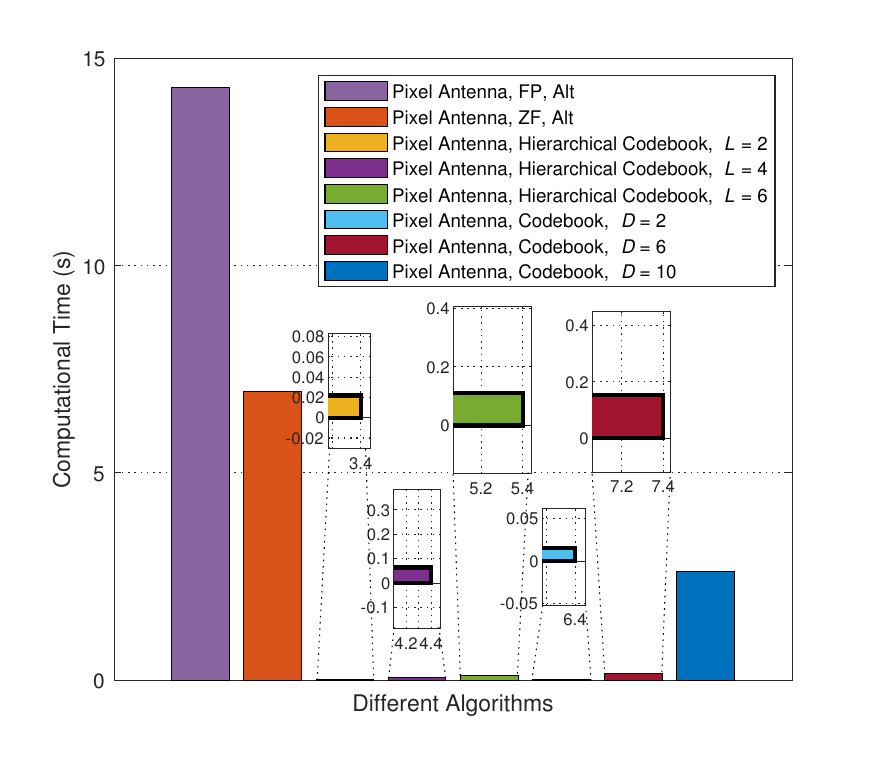}}
\subfigure[$N = U = 4$]{\includegraphics[width=0.485\textwidth]{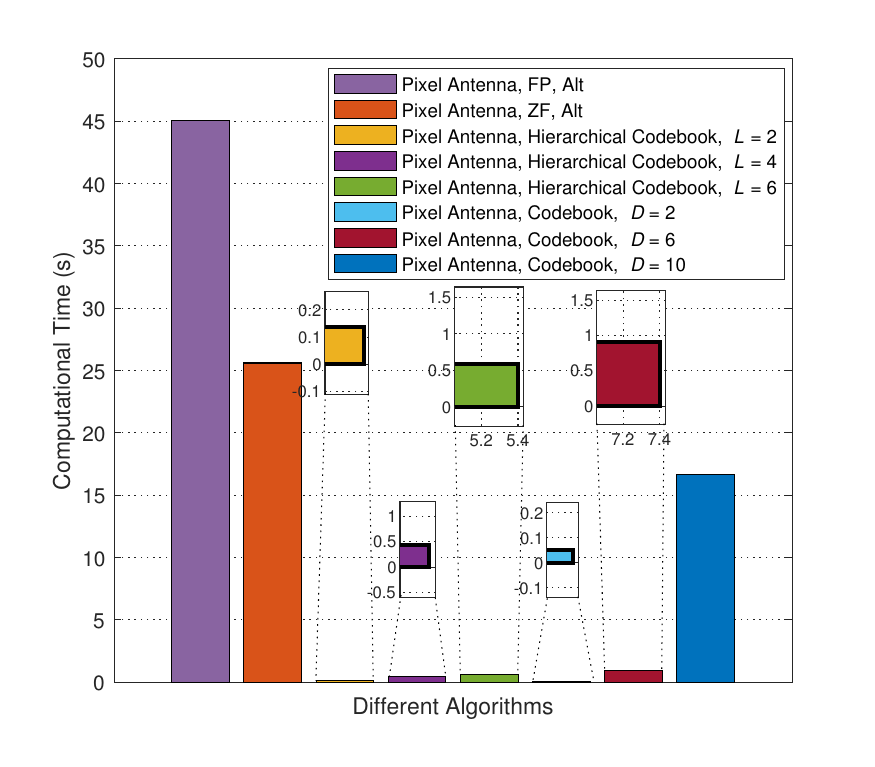}}
\caption{Average computational time of different algorithms for joint precoding
and antenna coding design.}
\label{fig:cputime} 
\end{figure*}

We also evaluate the computational time of the different algorithms
for the joint precoding and antenna coding design. Fig. \ref{fig:cputime}
illustrates the average computational time of the different algorithms
including ``FP, Alt'', ``ZF, Alt'', ``Codebook'', and ``Hierarchical Codebook'',
from which we have the following observations.

\textit{First}, combining the performance results in Fig. \ref{fig:SR_P}
and the computational time results in Fig. \ref{fig:cputime}, the
``FP, Alt'' algorithm achieves the best performance at the cost
of requiring the highest computational time among all algorithms.
This observation implies that while the ``FP, Alt'' algorithm has
satisfactory performance benefiting from rigorous mathematical derivations,
it may not be practical for real-world applications.

\textit{Second}, the codebook-based algorithm is much more computationally
efficient in terms of the computational time than “FP, Alt'' and
“ZF, Alt” algorithms. For example, when $N=U=4$, the codebook-based
algorithm can achieve around 97\% of the performance achieved by two
alternating algorithms at the cost of only 38\% of the computational
time of the ``FP, Alt'' algorithm and 65\% of the computational
time of the ``ZF, Alt'' algorithm. In addition, the codebook-based
algorithm with $D=6$ achieves around 91\% ($N=U=2$) and 88\% ($N=U=4$)
of the performance using the ``FP, Alt'' algorithm, while the former
requires only 1\% ($N=U=2$) and 2\% ($N=U=4$) of the computational
time of the latter.

\textit{Third}, the hierarchical codebook-based algoritm can be much
more computationally efficient than the codebook-based design with
proper numbers of layers. For example, when SNR $\le$ 15 dB, the
hierarchical codebook-based design with $L=6$ achieves almost the
same sum rate performance as the codebook-based design with $D=6$,
while the former saves 47\% of the computational time with $N=U=2$
and 37\% of the computational time with $N=U=4$. These results demonstrate
the superior computation efficiency of the proposed hierarchical codebook-based
algorithm.

Overall, the superior performance achieved by the ``FP, Alt'' algorithm is achieved at the cost of high computational complexity; with proper codebook sizes and layers, the ``Codebook'' and ``Hierarchical Codebook'' algorithms can achieve 
satisfactory performance with significantly reduced computational complexity.

\section{Conclusion}

\label{sec:conclusion}

In this work, we study the antenna coding design based on pixel antennas
for MU-MISO systems. Specifically, leveraging the fractional programming
theory, we first propose to alternatively optimize the antenna coder
using SEBO and transmit precoder using fractional programming to maximize
the sum rate performance. On top of this, to reduce the computational
complexity, we propose and develop a heuristic algorithm for the codebook
based antenna coding design. To further improve the computation efficiency,
we further propose a hierarchical codebook-based antenna coding design
based on a hierarchical search that achieves better performance-complexity
trade-off.

Simulation results show that when pixel antennas are deployed at the
user to replace conventional antennas with fixed configurations and
radiation patterns, the sum rate performance can be significantly
improved, e.g. doubled at SNR = 10 dB in an MU-MISO system including two conventional transmit antennas and two users with each having one pixel antenna. This performance enhancement comes from the reconfigurablity and adaptivity provided by pixel antennas
with optimized antenna coding, which achieves a significant channel
gain with better interference management. In addition, the proposed
(hierarchical) codebook-based algorithms can significantly reduce
the computational time while maintaining a satisfactory sum rate performance,
compared to the alternating algorithm using SEBO. Overall, this work
provides a low-complexity and high-performance antenna coding technique
empowered by pixel antennas for future multi-user transmission.

\vspace{-0.2 cm}

\bibliographystyle{IEEEtran}
\bibliography{refs}

\end{document}